\documentclass[12pt]{article}
\usepackage{epsfig}
\usepackage{amssymb}
\usepackage{amsmath}
\usepackage{amsfonts}
\usepackage{graphicx}
\usepackage{mathrsfs}
\usepackage{mathabx}
\DeclareMathAlphabet{\mathscrbf}{OMS}{mdugm}{b}{n}
\usepackage[dvips]{color}
\usepackage{multirow}
\usepackage{calc}
\usepackage{accents}

\makeatletter
\newcommand{\shorteq}{\mathrel{\mkern0.2mu\mathpalette\shorteq@\relax\mkern0.2mu}}
\newcommand{\shorteq@}[2]{\scalebox{0.5}[1]{$\m@th#1=$}}

\newcommand{\longeq}[1]{\mathrel{\mathpalette\longeq@{#1}}}
\newcommand{\longeq@}[2]{%
  \begingroup
  \sbox\z@{$\m@th#1=$}%
  \ifdim#2<\wd\z@
    \resizebox{#2}{\height}{\box\z@}%
  \else
    \ifdim#2<3\wd\z@
      \hbox to #2{$\m@th#1=\hss=\hss=\hss=$}%
    \else
      \hbox to #2{$\m@th#1=\cleaders\hbox to 0.2\wd\z@{\hss$#1=$\hss}\hfil=$}%
    \fi
  \fi
  \endgroup
}
\makeatother

\newcommand{\R}{\mathbb{R}}
\newcommand{\C}{\mathbb{C}}

\newcommand{\fR}{\mathfrak{R}}

\newcommand{\cP}{\mathcal{P}}

\newcommand{\cT}{\mathcal{T}}
\newcommand{\cU}{\mathcal{U}}

\newcommand{\be}{\begin{equation}}
\newcommand{\ee}{\end{equation}}
\newcommand{\bea}{\begin{eqnarray}}
\newcommand{\eea}{\end{eqnarray}}
\newcommand{\nn}{\nonumber}
\newcommand{\kt}{\rangle}
\newcommand{\br}{\langle}

\newcommand{\ed}{\end{document}}

\newcommand{\bi}{\begin{itemize}}
\newcommand{\ei}{\end{itemize}}

\newcommand{\bce}{\begin{center}}
\newcommand{\ece}{\end{center}}

\newcommand{\sD}{\mathscr{D}}

\newcommand{\sH}{\mathscr{H}}

\newcommand{\sR}{\mathscr{R}}

\newcommand{\RE}{{\rm Re}}
\newcommand{\IM}{{\rm Im}}

\newcommand{\for}{{\mbox{\rm for}}}

\begin{document}

%\title{Exact Solution of the Scattering Problem for TE and TM Waves Scattered by a Class of Diffraction Gratings}

\title{Does Born Rule Imply Unitarity of Time Evolution in Quantum Mechanics?}

\author{Ali~Mostafazadeh\thanks{E-mail address: amostafazadeh@ku.edu.tr}\\[6pt]
Departments of Mathematics and Physics, Ko\c{c} University,\\  34450 Sar{\i}yer,
Istanbul, T\"urkiye}

\date{ }
\maketitle

\begin{abstract} The Born rule for computing probabilities of the outcomes of measurements is an indispensable ingredient of quantum mechanics. The standard textbook description of this rule gives the impression that it implies the unitarity of time evolution. This view relies on the argument that unless the dynamics is unitary, the probabilities of finding all possible outcomes of a measurement do not add up to 1, i.e., the total probability is not conserved. We show that this argument is flawed, and that the general expression for the Born rule ensures the conservation of total probabilities even when the dynamics of a quantum system is not unitary. This applies to the dynamics of ensembles of quantum systems in both pure and mixed states. We discuss the status of the local conservation of probabilities and the arguments against the plausibility of non-unitary time evolutions that are based on the identification of the Hamiltonian operator with the energy observable.

\end{abstract}

\section{Introduction} 

The probabilistic nature of quantum mechanics is the essential feature that distinguishes it from classical mechanics. It was originally formulated by Max Born in 1926 \cite{Born-1926}, and later made into one of the fundamental postulates of quantum mechanics by von Neumann in 1932 \cite{von-Neumann}. This is known as the measurement postulate and forms the basis of the predictive power of quantum mechanics. 

To give the statement of the measurement postulate, we recall the postulates identifying the states and observables of quantum systems. According to these, the states of a quantum mechanical system are determined by nonzero elements $\psi$ of a complex Hilbert space $\sH$, which are called state vectors, and its observables are represented by Hermitian operators\footnote{Throughout this article we use the term ``Hermitian'' to mean ``self-adjoint'' as is done by von~Neumann in Ref.~\cite{von-Neumann}.} acting in $\sH$. 

It is important to note that for each nonzero complex number $c$, the state vector $c\psi$ identifies the same state as $\psi$. This implies that there is a one-to-one correspondence between the states of a system and the one-dimensional subspaces $\sR$ of its Hilbert space $\sH$. These are sometimes called the rays in the Hilbert space. Because we can use every state vector $\psi$ belonging to a state $\sR$ to express it in the form $\{\,c\psi\,|\,c\in\C\}$, $\psi$ determines $\sR$ in a unique manner. The converse is clearly not true, because $\sR$ includes infinity many state vectors.

The postulate of quantum mechanics that outlines its dynamical features states that the evolution of states of a quantum system is determined by the action of a time-dependent unitary operator $\widehat U(t,t_0)$ acting in its Hilbert space, where $t$ stands for time, and $t_0$ is the initial time. This means that the state of the system at time $t$ is given by the state vector
	\be
	\psi(t):=\widehat U(t,t_0)\psi_0,
	\label{ev-op} 
	\ee 
where $\psi_0$ is any state vector that determines the state of the system at $t_0$. $\widehat U(t,t_0)$ is called the evolution operator.

Assuming that $\psi(t)$ is differentiable, we can easily show that it satisfies the time-dependent Schr\"odinger equation,
	\be
	i\hbar\frac{d}{dt} \psi(t)=\widehat H(t)\psi(t),
	\label{sch-eq}
	\ee
where $\widehat H(t):=i\hbar[\frac{d}{dt}\widehat U(t,t_0)]\widehat U(t,t_0)^{-1}$ is the Hamiltonian of the system. Demanding that $\widehat U(t,t_0)$ is unitary implies that $\widehat H(t)$ is a Hermitian operator.

In practice, quantum systems are identified with the corresponding Hamiltonian operators, and the evolution operator $\widehat U(t,t_0)$ is treated as a derived quantity. Substituting \eqref{ev-op} in \eqref{sch-eq}, we find that the latter satisfies
	\be
	i\hbar\frac{d}{dt}\,\widehat U(t,t_0)=\widehat H(t)\,\widehat U(t,t_0),
	\qquad\qquad \widehat U(t_0,t_0)=\widehat I,
	\nn%\label{ev-op-eq}
	\ee
where $\widehat I$ stands for the identity operator acting in $\sH$. If the Hamiltonian does not depend on time, its Hermiticity ensures the existence, uniqueness, and unitarity of the evolution operator $\widehat U(t,t_0)$ for all $t_0$ and $t$, \cite{reed-simon}. If the Hilbert space is finite-dimensional, $\widehat U(t,t_0)$ exists for all $t_0$ and $t$ and is unique regardless of whether the Hamiltonian is time-dependent or Hermitian \cite{reed-simon}. In cases for which $\widehat H(t)$ is non-Hermitian, $\widehat U(t,t_0)$ is not a unitary operator, and the system undergoes a non-unitary time evolution.

Relaxing the condition that the Hamiltonians must be Hermitian while insisting on the validity of the other postulates of quantum mechanics encounters the difficulty that non-Hermitian Hamiltonians generate non-unitary time evolutions which in light of the standard description of the Born rule for computing probabilities of outcomes of measurements \cite{griffiths-QM,sakurai} seems to imply that these probabilities do not add up to 1. Specifically, their sum changes in time, i.e., total probabilities are not conserved. This line of thought has motivated the search for replacements of the principle of conservation of total probabilities for systems defined by non-Hermitian Hamiltonians \cite{Torosov,Kivela}.

An indirect evidence against the argument posing unitarity of time evolution as a consequence of the Born rule is that it conflicts with the fact that measurements in quantum mechanics are instantaneous processes which should not have any bearing on the dynamical features of the system; how can the extract of information about the state of a system at a single instant of time can restrict the nature of its dynamical evolution during the time periods in which no such information is extracted?

The main purpose of the present article is to show that the Born rule, written in a way that makes the probabilities invariant under the general scalings of the state vectors, $\psi\to c\psi$, by nonzero complex numbers $c$, can be consistently applied for quantum systems regardless of whether they undergo a unitary or non-unitary dynamics. 

The outline of the remainder of this article is as follows. In Sec.~2, we provide a detailed account of the standard argument that presents unitarity as an inevitable consequence of the Born rule and conservation of total probabilities. Here we reveal the flaw in this argument and establish the consistency of the general expression for the Born rule and the unitarity of time evolution both for individual quantum systems and statistical ensembles of such systems. In Sec.~3, we discuss the nature and implications of local conservation of probabilities. In Sec.~4, we examine the arguments against non-unitary time evolutions that are based on the requirement that the Hamiltonian operator represents the energy observable. In Sec.~5, we present our concluding remarks.

\section{Born rule and conservation of total probabilities}

Suppose for simplicity that $\sH$ is a finite-dimensional Hilbert space, and consider a Hermitian operator $\widehat O$ representing an observable of the system that has a non-degenerate spectrum. Then, $\sH$ admits an orthonormal basis $\{\psi_n\}_{n=1}^N$ consisting of eigenvectors of $\widehat O$, where $N$ is the dimension of $\sH$. This means that we can expand any element $\psi$ of $\sH$ as a linear combination of $\psi_n$'s; there are unique complex numbers $c_1, c_2,\cdots,c_N$ such that
	\be
	\psi=c_1\psi_1+c_2\psi_2+\cdots+c_N\psi_N.
	\label{expand}
	\ee
Because $\{\psi_n\}_{n=1}^N$ is orthonormal, 
	\be
	\br\psi_n|\psi_m\kt=
	\delta_{mn}:=
	\left\{\begin{array}{ccc}
	1 &\for & m=n,\\
	0 &\for &m\neq 0.
	\end{array}\right.
	\label{orthonormal}
	\ee
We can use this relation to compute the inner product of both sides of \eqref{expand} with $\psi_n$, which gives 
	\be
	c_n=\br\psi_n|\psi\kt.
	\label{cn=}
	\ee
	
Another consequence of \eqref{expand} and \eqref{orthonormal} is the Pythagorean identity: 		\be
	\parallel\!\psi\!\parallel^2%\:=\br\psi|\psi\kt
	=|c_1|^2+|c_2|^2+\cdots+|c_n|^2,
	\label{pythagore}
	\ee 
where $\parallel\!\psi\!\parallel$ stands for the norm of $\psi$, i.e., $\parallel\!\psi\!\parallel:=\sqrt{\br\psi|\psi\kt}$. Substituting  \eqref{cn=} in \eqref{pythagore}, we have
	\be
	\parallel\!\psi\!\parallel^2\:=\sum_{n=1}^N|\br\psi_n|\psi\kt|^2.
	\label{norm2}
	\ee

The measurement postulate of quantum mechanics states that the outcome of a measurement of $\widehat O$ is one of its eigenvalues, and that we can only compute the probabilities of finding these eigenvalues. Specifically, if the state of the system immediately before the measurement is given by the state vector $\psi$, the probability ${\rm Prob}_n$ of finding the eigenvalue $\omega_n$ corresponding to the eigenvector $\psi_n$ is given by the Born rule \cite{sakurai,cohen-tannoudji}:
	\be
	{\rm Prob}_n=|c_n|^2=|\br\psi_n|\psi\kt|^2.
	\label{BR}
	\ee
Because sum of the probabilities of all possible outcomes of any measurement must be equal to $1$,  \eqref{norm2}  and \eqref{BR} imply
	\be
	\parallel\!\psi\!\parallel^2\:=\sum_{n=1}^N {\rm Prob}_n=1.
	\label{eq1}
	\ee
This calculation shows that the Born rule, as expressed by \eqref{BR}, applies only if $\psi$ is a normalized state vector, i.e., $\parallel\!\psi\!\parallel=1$, \cite{sakurai,cohen-tannoudji}. 

If $\psi$ describes the state of the system at time $t$, i.e., it is a solution of the Schr\"odineger equation \eqref{sch-eq}, Eq.~\eqref{eq1} becomes
	\be
	\parallel\!\psi(t)\!\parallel^2\:=\sum_{n=1}^N {\rm Prob}_n(t)=1,
	\label{eq2}
	\ee
where ${\rm Prob}_n(t)$ is the probability of finding $\omega_n$ if we measure $\widehat O$ at time $t$. According to \eqref{eq2}, $\psi(t)$ remains normalized for all $t$. Substituting \eqref{ev-op} in \eqref{eq2}, we have
	\be
	\parallel\!\widehat U(t,t_0)\psi_0\!\parallel^2=1=\parallel\!\psi_0\!\parallel^2.
	\label{eq3}
	\ee
This relation shows that $\widehat U(t,t_0)$ does not change the norm of the initial state vector $\psi_0$ provided that the latter is normalized. It is easy to show that this is also true when $\psi_0$ is a non-normalized element of $\sH$.\footnote{Suppose that $\psi_0$ is an arbitrary element of $\sH$. If $\psi_0=0$, $\widehat U(t,t_0)\psi_0=0$, and $\parallel\!\widehat U(t,t_0)\psi_0\!\parallel=0=\parallel\!\psi_0\!\parallel$. If $\psi_0\neq 0$, we introduce $c_0:=\parallel\!\psi_0\!\parallel$ and $\hat\psi_0:=c_0^{-1}\psi_0$. Because $\hat\psi_0$ is normalized, we can apply \eqref{eq3} and conclude that $\parallel\!\widehat U(t,t_0)\hat\psi_0\!\parallel^2=1$. This equation together with $\psi_0=c_0\hat\psi_0$ and the fact that $\widehat U(t,t_0)$ is a linear operator imply $\parallel\!\widehat U(t,t_0)\psi_0\!\parallel^2=
\parallel c_0\,\widehat U(t,t_0)\hat\psi_0\!\parallel^2=
|c_0|^2\parallel\!\widehat U(t,t_0)\hat\psi_0\!\parallel^2=|c_0|^2=\parallel\!\psi_0\!\parallel^2$.} Therefore $\widehat U(t,t_0)$ leaves the norms of all elements of $\sH$ unchanged, i.e., it is a unitary operator. This is the content of the argument supporting the claim that unitarity of time evolution is a necessary condition for the conservation of total probabilities \cite{griffiths-QM,Landau-Lifshitz-QM}. 

The main problem with this argument is that it relies on Eq.~\eqref{BR} which holds only if $\psi$ is normalized. If the system's time evolution is not unitary, we are not allowed to use Eq.~\eqref{BR} for calculating ${\rm Prob}_n$ at times $t>t_0$ even if we assume that $\psi_0$ is normalized. Therefore, the hypothesis of this argument is false which renders it inconclusive. This raises the question if there is a way to compute ${\rm Prob}_n$ without assuming that the state vector $\psi$ is normalized. We can do this simply by expressing the Born rule in the form \cite{Fock}:
	\be
	{\rm Prob}_n=\frac{|c_n|^2}{|c_1|^2+|c_2|^2+\cdots+|c_n|^2}=
	\frac{|\br\psi_n|\psi\kt|^2}{\br\psi|\psi\kt},
	\label{BR-2}
	\ee
which clearly ensures that
	\be
	\sum_{n=1}^N {\rm Prob}_n=1.
	\label{total}
	\ee
Equation~\eqref{BR-2} does not restrict the choice of $\psi$ except that $\psi\neq 0$. For normalized $\psi$, it coincides with \eqref{BR}.  More importantly, it shows that ${\rm Prob}_n$ is manifestly invariant under the scalings, $\psi\to c\psi$, of the state vector $\psi$ for $c\neq 0$. This means that ${\rm Prob}_n$ depends on the state of the system $\sR:=\{c\psi|c\in\C\}$ and not on the choice of the corresponding state vectors.

For an evolving state vector $\psi(t)$, \eqref{BR-2} and \eqref{total} take the forms
	\begin{align}
	&{\rm Prob}_n(t)=\frac{|\br\psi_n|\psi(t)\kt|^2}{\br\psi(t)|\psi(t)\kt},
	\label{BR-2-t}\\
	&\sum_{n=1}^N {\rm Prob}_n(t)=1.
	\label{total-t}
	\end{align}
Again, the Born rule \eqref{BR-2-t} holds if $\psi(t)\neq 0$ and $\{\psi_n\}_{n=1}^N$ is an orthonormal basis of $\sH$. The latter implies the completeness relation, 
	\be 
	\sum_{n=1}^N\widehat\Pi_n=\widehat I,
	\label{complete}
	\ee
where 
	\be
	\widehat\Pi_n:=|\psi_n\kt\br\psi_n|,
	\label{Pi-n=}
	\ee
and we have employed Dirac's bra-ket notation. It is easy to see that $\widehat\Pi_n$ is the projection operator onto the eigenspace $\sH_n$ associated with $\omega_n$, i.e.,
	\be
	\sH_n:=\{\,\phi\in\sH\,|\,\widehat O\phi=\omega_n\phi\,\}.
	\ee
We can express the Born rule \eqref{BR-2-t} in terms of $\widehat\Pi_n$. This follows from the fact that 
	\[|\br\psi_n|\psi(t)\kt|^2=\br\psi(t)|\psi_n\kt\br\psi_n|\psi(t)\kt=
	\br\psi(t)|\widehat\Pi_n\psi(t)\kt.\]
Substituting this relation in \eqref{BR-2-t}, we find
	\begin{align}
	&{\rm Prob}_n(t)=\frac{\displaystyle \br\psi(t)|\widehat
	\Pi_n\psi(t)\kt}{\br\psi(t)|\psi(t)\kt}.
	\label{BR-3-t}
	\end{align}
In view of this equation, we can identify \eqref{total-t} as a consequence of the completeness relation  \eqref{complete}. 

Because the above analysis does not involve the use of the Schr\"odinger equation~\eqref{sch-eq}, \eqref{total-t} holds true regardless of whether the Hamiltonian of the system is Hermitian or not. This shows that the Born rule as given by \eqref{BR-2-t} applies to systems with unitary as well as non-unitary time evolutions.\footnote{The only condition it imposes on the dynamics of the system is that $\psi(t)\neq 0$ for all $t$. This is equivalent to demanding that the evolution operator be invertible which follows from the uniqueness of the solution of the Schr\"odinger equation.}

The expression of the Born rule provided by \eqref{BR-3-t} applies also for cases where the observable $\widehat O$ has a degenerate spectrum. In this case, the projection operator that maps $\sH$ onto $\sH_n$ takes the form
	\be
	\widehat\Pi_n:=\sum_{a=1}^{d_n}|\psi_{n,a}\kt\br\psi_{n,a}|,
	\label{eq-z1}
	\ee
where $d_n$ is the degree of degeneracy of $\omega_n$, and $\{\psi_{n,a}\}_{a=1}^{d_n}$ is an orthonormal basis of $\sH_n$. It is easy to see that 
	\begin{align}
	&\widehat\Pi_m\widehat\Pi_n=\delta_{mn}\widehat\Pi_n,
	\label{ortho}\\
	&\sum_{n=1}^d\widehat\Pi_n=\widehat I,
	\label{complete-4}
	\end{align}
where $d$ is the number of distinct eigenvalues of $\widehat O$. Eqs.~\eqref{BR-3-t} and \eqref{complete-4} imply 
	\be
	\sum_{n=1}^d {\rm Prob}_n(t)=1.
	\label{conserve}
	\ee
This shows that the Born rule \eqref{BR-3-t} for observables with a degenerate spectrum also ensures the conservation of total probabilities without imposing any restriction on the Hamiltonian and the evolution operator of the system.

It is not difficult to generalize the above argument to statistical ensembles of quantum systems whose states are given by density operators $\widehat\rho$, \cite{sakurai}. For a system whose dynamics is determined by a possibly non-Hermitian Hamiltonian operator, the density operators evolve in time according to
	\be
	\widehat\rho(t)=\widehat U(t,t_0)\,\widehat\rho_0\,\widehat U(t,t_0)^\dagger,
	\label{rho-t}
	\ee
where $\widehat\rho_0$ is the density operator describing the state of the ensemble at initial time $t_0$.

To derive the Born rule for an statistical ensemble of quantum systems in a state given by $\widehat\rho(t)$, first we confine our attention to cases where $\widehat\rho_0$ corresponds to a pure state. Then, all the members of the ensemble are in a state given by a state vector $\psi_0\in\sH$, and the initial state of the ensemble is given by $\widehat\rho_0=|\psi_0\kt\br\psi_0|$. In view of \eqref{ev-op} and \eqref{rho-t}, this shows that the density operator representing the state of the ensemble at time $t$ has the form
	\be
	\widehat\rho(t)=|\psi(t)\kt\br\psi(t)|.
	\label{rho=}
	\ee
It is also clear that in this case the probability of finding $\omega_n$ by measuring $\widehat O$ is given by \eqref{BR-3-t}.  

Next, we recall that the trace of a linear operator $\widehat L$ is given by
	\be
	{\rm tr}[\widehat L]:=\sum_{m=1}^N\br\phi_m|\widehat L\phi_m\kt,
	\label{trace}
	\ee
where $\{\phi_m\}_{m=1}^N$ is an arbitrary orthonormal basis  of $\sH$. Using the orthonormal basis consisting of the eigenvectors $\psi_{n,a}$ of $\widehat O$ to compute ${\rm tr}[\widehat \Pi_n\widehat\rho(t)]$, we find
	\begin{align}
	{\rm tr}[\widehat \Pi_n\widehat\rho(t)]&=\sum_{m=1}^d\sum_{a=1}^{d_m}
	\br\psi_{m,a}|\widehat\Pi_n\psi(t)\kt\br\psi(t)|\psi_{m,a}\kt\nn\\
	&=\sum_{m=1}^d\sum_{a=1}^{d_m}
	\br \widehat\Pi_n\psi_{m,a}|\psi(t)\kt\br\psi(t)|\psi_{m,a}\kt\nn\\
	&=\sum_{a=1}^{d_n}
	\br\psi_{n,a}|\psi(t)\kt\br\psi(t)|\psi_{n,a}\kt\nn\\
	&=\sum_{a=1}^{d_n}
	\br\psi(t)|\psi_{n,a}\kt	\br\psi_{n,a}|\psi(t)\kt\nn\\
	&=\br\psi(t)|\widehat\Pi_n\psi(t)\kt,
	\label{eq-z2}\\
	{\rm tr}[\widehat\rho(t)]&=
	\sum_{n=1}^d {\rm tr}[\widehat \Pi_n\widehat\rho(t)]=\br\psi(t)|\psi(t)\kt,
	\label{eq-z3}
	\end{align}
where we have made use of the Hermiticity of $\widehat\Pi_n$, and the identity, $\widehat\Pi_n \psi_{m,a}=\delta_{mn} \psi_{n,a}$, which follows from \eqref{eq-z1} and  \eqref{complete-4}. In view of \eqref{eq-z2} and \eqref{eq-z3}, we can express the Born rule \eqref{BR-3-t} in the form
	\be
	{\rm Prob}_n(t)=\frac{{\rm tr}[\widehat \Pi_n\widehat\rho(t)]}{{\rm tr}[\widehat\rho(t)]}.
	\label{BR-5-t}
	\ee
 {This is the form of the Born rule that applies to pure states.} 

 {According to \eqref{eq-z3}, the evolution operator is unitary if and only if for all pure states the trace of the corresponding density operator does not depend on time, i.e., the dynamics is trace-preserving. If the time-evolution operator is not unitary, ${\rm tr}[\widehat\rho(t)]$ changes in time. In particular, it cannot have unit trace for all $t$. This observation shows that the standard assumption that the density operators must have unit trace is violated for non-unitary evolutions. Therefore, to examine the Born rule for a mixed state undergoing a non-unitary evolution, we must relax this condition. We do this by identifying density operators with positive operators (Hermitian operator having nonnegative eigenvalues) $\widehat\rho:\sH\to\sH$ whose trace is nonzero.}

 {We can express density operators in terms of their spectral resolution,
	\begin{align}
	&\widehat\rho=\sum_{a=1}^{N} r_a\,\widehat\rho_a,
	&&\widehat\rho_a:=|\phi_a\kt\br\phi_a|,
	\label{mixed=}
	\end{align}
where $r_1,r_2,\cdots,r_N$ are (possibly repeated) eigenvalues of $\widehat\rho$, and $\{\phi_a\}_{a=1}^N$ is an orthonormal basis of $\sH$ consisting of the eigenvectors of $\widehat\rho$. We can use the latter condition to deduce the identities
	\begin{align}
	&{\rm tr}(\widehat\rho_a)=1,
	&&{\rm tr}(\widehat\rho)=\sum_{a=1}^N r_a.
	\label{traces}
	\end{align}
Since $\widehat\rho$ is a positive operator with a nonzero trace, $r_a\geq 0$ and ${\rm tr}(\widehat\rho)>0$. In particular, we can view 
	\be
	p_a:=\frac{r_a}{r_1+r_2+\cdots+r_N}=\frac{r_a}{{\rm tr}(\widehat\rho)}
	\label{pa=}
	\ee
as the probability of the ensemble to be in the pure state $\sR_a$ given by $\phi_a$.}

 {Now, consider a measurement of the observable $\widehat O$. If the ensemble is in the pure state $\sR_a$, we can use \eqref{BR-5-t} and  \eqref{traces} to compute the probability of finding $\omega_n$. The result is ${\rm tr}[\widehat \Pi_n\widehat\rho_a]$. This implies that the probability of finding $\omega_n$ by measuring the observable $\widehat O$ when the ensemble is in the mixed state given by $\widehat\rho$ is the sum of the products of $p_a$'s and ${\rm tr}[\widehat \Pi_n\widehat\rho_a]$'s, i.e., 
	\be
	{\rm Prob}_n=\sum_{a=1}^N p_a 
	{\rm tr}[\widehat \Pi_n\widehat\rho_a]=\sum_{a=1}^N r_a 
	\frac{{\rm tr}[\widehat \Pi_n\widehat\rho_a]}{{\rm tr}(\widehat\rho)}=
	\frac{1}{{\rm tr}[\widehat\rho]}
	{\rm tr}\Big[\sum_{a=1}^Nr_n\widehat\rho_s\Big]	
	=
	\frac{{\rm tr}[\widehat \Pi_n\widehat\rho]}{{\rm tr}[\widehat\rho]},
	\label{BR-15-t}
	\ee
where we have used \eqref{mixed=}, \eqref{pa=}, and the fact that trace is a linear mapping of operators to complex numbers.}

 {If the measurement of $\widehat O$ takes place at time $t$, the ensemble is in the mixed state given by a density operator $\widehat\rho(t)$ of the form \eqref{rho-t}, and \eqref{BR-15-t} coincides with \eqref{BR-5-t}. This shows that this equation holds true for both pure and mixed states.} Again, we can use \eqref{complete-4} to show that \eqref{BR-5-t} implies \eqref{conserve}. Therefore, the total probabilities are conserved regardless of whether the system undergoes a unitary or non-unitary evolution.

 {An immediate consequence of the Born rule \eqref{BR-5-t} for the (mixed) states is its invariance under the scalings, $\widehat\rho\to c\,\widehat\rho$, where $c\in\R^+$. This suggests identifying (pure and mixed) states with the ``rays'' of density operators, $\fR:=\{ c\,\widehat\rho\,|\,c\in\R^+\}$. We can use any element of $\fR$ to compute the probabilities of outcomes of measurements when the ensemble is in the state $\fR$. In particular, we can use the ``normalized density operator,''
	\be
	\widehat{\hat\rho}:=[{\rm tr}(\widehat\rho)]^{-1}\widehat\rho,
	\label{nor-rho}
	\ee
which belongs to $\fR$ and has unit trace.}

 {When the system undergoes a unitary time evolution, an initial normalized density operator $\widehat{\hat\rho}_0$ remains normalized at all times. To see this, we use \eqref{rho-t} to show that
	\begin{align}
	{\rm tr}\big[\widehat{\rho(t)}\big]&={\rm tr}\big[\widehat U(t,t_0)\,\widehat{\hat\rho}_0\,
	\widehat U(t,t_0)^\dagger\big]\nn\\
	&={\rm tr}\big[\widehat U(t,t_0)^\dagger\widehat U(t,t_0)\,\widehat{\hat\rho}_0
	\big]\nn\\
	&={\rm tr}\big[\,\widehat{\hat\rho}_0\big]=1, 
	\label{new-id}
	\end{align}
where we have also employed the identity, ${\rm tr}(\widehat A\widehat B)={\rm tr}(\widehat B\widehat A)$, which holds for any pair of linear operators $\widehat A$ and $\widehat B$. Equation~\eqref{new-id} allows us to express the Born rule \eqref{BR-5-t} in the form 
	\be
	{\rm Prob}_n(t)={\rm tr}[\widehat \Pi_n\widehat{\rho}(t)].
	\label{BR-5-t-normalized}
	\ee
This is the standard formula for computing probabilities of mixed states, but it only applies to systems undergoing unitary evolutions. We can use it for systems undergoing non-unitary evolutions provided that we normalize the evolving density operator $\widehat{\rho}(t)$, i.e., write  \eqref{BR-5-t} as
	\be
	{\rm Prob}_n(t)={\rm tr}[\widehat \Pi_n\widehat{\hat\rho}(t)].
	\label{BR-5-t-normalized-2}
	\ee
where $\widehat{\hat\rho}(t):={\rm tr}[\widehat{\rho}(t)]^{-1}\widehat{\rho}(t)$.}
	
 Next, suppose that there are real and positive numbers $c_1$ and $c_2$, and density operators $\widehat\rho_1$ and $\widehat\rho_1$, such that $\widehat\rho=c_1\widehat\rho_1+c_2\widehat\rho_1$. Let ${\rm Prob}^{(1)}_n$ and ${\rm Prob}^{(2)}_n$ stand for the probabilities of finding $\omega_n$ as a result of a measurement of $\widehat O$ when the ensemble is in the state given by $\widehat\rho_1$ and $\widehat\rho_2$, respectively. If $\widehat\rho_1$, $\widehat\rho_2$, and $\widehat\rho$ are normalized, which happens when $c_1+c_2=1$, we can use  \eqref{BR-5-t-normalized} to show that
	\be
	{\rm Prob}_n=c_1\,{\rm Prob}^{(1)}_n+c_2\,{\rm Prob}^{(1)}_n.
	\label{linearity-txt}
	\ee
When either of $\widehat\rho_1$, $\widehat\rho_2$, and $\widehat\rho$ is not normalized, we must use \eqref{BR-15-t} to compute ${\rm Prob}^{(1)}_n$, ${\rm Prob}^{(2)}_n$, and ${\rm Prob}_n$, and it seems that the linearity property \eqref{linearity-txt} is violated.  In Appendix~A, we derive the generalization of \eqref{linearity-txt} for situations where $\widehat\rho_1$, $\widehat\rho_2$, and $\widehat\rho$ need not be normalized. The result is that the probabilities of the ensemble to be in the states given by $\widehat\rho_1$ and $\widehat\rho_2$ do not correspond to $c_1$ and $c_2$. They are instead given by 
	\begin{align}
	&\hat c_1:=\frac{c_1\,{\rm tr}\big(\widehat\rho_1\big)}{
	c_1 {\rm tr}\big(\widehat\rho_1\big)+c_2 {\rm tr}\big(\widehat\rho_2\big)},
	&&\hat c_2:=\frac{c_2\,{\rm tr}\big(\widehat\rho_2\big)}{
	c_1 {\rm tr}\big(\widehat\rho_1\big)+c_2 {\rm tr}\big(\widehat\rho_2\big)},
	\nn%\label{appB-4t}
	\end{align}
respectively, and \eqref{linearity-txt} generalizes to
	\be
	{\rm Prob}_n=\hat c_1\,{\rm Prob}^{(1)}_n+\hat c_2\,{\rm Prob}^{(1)}_n,
	%\label{linearity-txt2}
	\ee
which is consistent with the basic rules of probability theory.	

\section{Local conservation of probabilities}

Consider a quantum system obtained by the standard canonical quantization of a particle moving in a one-dimensional Euclidean space. The Hilbert space of this system is the space $L^2(\R)$ of square-integrable functions, $\psi:\R^3\to\C$, and its Hamiltonian has the form
	\be
	\widehat H=\frac{\widehat  P^2}{2m}+v(\widehat X),
	\label{H=}
	\ee
where $m$ is the mass of the particle, $\widehat X$ and $\widehat P$ are respectively the standard position and momentum operators, and $v$ is a time-independent potential.\footnote{The inner product in $L^2(\R)$ is given by $\br\chi_1|\chi_2\kt:=\int_{-\infty}^\infty dx\, \chi_1(x)^*\chi_2(x)$, and the position and momentum operators are respectively defined by $\big(\widehat X\phi\big)(x):=x \phi(x)$ and $\big(\widehat P\phi\big)(x):=-i\hbar\,\frac{d}{dx}\,\phi(x)$, where $x\in\R$ and $\phi$ belongs to a common domain $\sD$ of both of these operators. A suitable choice for $\sD$ is the set of functions $\phi$ with derivatives of all orders such that the products of its derivatives and any polynomial decay to zero as $x\to\pm\infty$. Both of these operators are Hermitian, their point spectra are empty, and their continuous spectra coincide with $\R$.} 

It is customary to express the Born rule applied to the position observable in the form \cite{griffiths-QM,sakurai}:
	\begin{align}
	{\rm Prob}_A(t):=\int_A dx\:|\br x|\psi(t)\kt|^2,
	\label{e3-1}
	\end{align}
where $A$ is a subset of $\R$, ${\rm Prob}_A(t)$ is the probability of finding the position of the particle in $A$, and $\psi(t)$ is the state vector determining the state of the particle at time $t$.\footnote{Strictly speaking $A$ is the union of a set of intervals of real numbers.} Again the condition that probabilities must add to 1 requires that 
	\be
	{\rm Prob}_\R(t)=1.
	\label{eqzz1}
	\ee 
In view of \eqref{e3-1}, this implies 
	\be 
	\br\psi(t)|\psi(t)\kt=\int_{-\infty}^\infty dx\:|\br x|\psi(t)\kt|^2=1.
	\label{normalization}
	\ee
This relation is sometimes taken as a justification for the unitarity of time evolution \cite{griffiths-QM}. This argument is flawed, because when the system undergoes a non-unitary evolution, e.g. when $v$ is a complex potential, one cannot employ \eqref{e3-1}. Instead, one must use the  generalization of \eqref{BR-2-t} given by
	\begin{align}
	{\rm Prob}_A(t):=\frac{1}{\parallel\psi(t)\parallel^2}\int_A dx\:|\br x|\psi(t)\kt|^2,
	\label{e3-2}
	\end{align}
which ensures \eqref{eqzz1}.  

We can express \eqref{e3-2} in the form
	\begin{align}
	{\rm Prob}_A(t):=\frac{\br\psi(t)|\widehat\Pi_A\psi(t)\kt}{\br\psi(t)|\psi(t)\kt},
	\label{e3-3}
	\end{align}
where
	\be
	\widehat\Pi_A:=\int_A dx~|x\kt\br x|,
	\label{e3-20}
	\ee
and $|x\kt$ stands for the Dirac delta function $\delta_x$ centered at $x$, i.e.,  $\br x'|x\kt=\delta_x(x'):=\delta(x-x')$. This in turn implies the completeness relation for $|x\kt\br x|$ which we can express as 
	\be
	\widehat\Pi_\R=\int_{-\infty}^\infty dx~|x\kt\br x|=\widehat I.
	\label{complex-7}
	\ee
We can view \eqref{eqzz1} as a consequence of \eqref{e3-3} and \eqref{complex-7}. 

A more important fact about the probabilities of localization of such a particle in the position space is their local conservation. This is a consequence of the continuity equation for the localization probabilities whose standard derivation assumes that the potential is real-valued. It is instructive to explore what happens if we try to arrive at a continuity equation for probabilities without making this assumption from the outset.  
 
According to \eqref{e3-2}, the probability density for the localization of the particle in the position space has the form
  	\be
	\varrho(x,t):=\frac{|\br x|\psi(t)\kt|^2}{\parallel\psi(t)\parallel^2}=
	\frac{|\psi(x,t)|^2}{\parallel\psi(t)\parallel^2},
	\label{rho=}
	\ee
where $\psi(x,t):=\br x|\psi(t)\kt$ is the position wave function which solves the Schr\"odinger equation \eqref{sch-eq} in the position representation,
	\be
	i\hbar\partial_t\psi(x,t)=
	-\frac{\hbar^2}{2m}\,\partial_x^2\psi(x,t)+v(x)\psi(x,t).
	\label{sch-eq-x}
	\ee
To derive the continuity equation underlying the local conservation of localization probabilities in position space, we proceed as follows. 
	\begin{enumerate}
	\item We use \eqref{H=} and the Hermiticity of $\widehat P$ to calculate 
	\begin{align}
	\frac{d}{dt}\parallel\!\psi(t)\!\parallel^2&= \frac{d}{dt}\,\br\psi(t)|\psi(t)\kt=
	2\,\RE\Big[\br\psi(t)|\frac{d}{dt}\psi(t)\kt\Big]\nn\\
	&=\frac{2}{\hbar}\,\IM\Big[\frac{1}{2m}\,\br\psi(t)|\widehat P^2\psi(t)\kt+\br\psi(t)|v(\widehat X)\psi(t)\kt\Big]
	\nn\\
	&=\frac{2}{\hbar}\,\IM\Big[\frac{1}{2m}\,\br\widehat P\psi(t)|\widehat P\psi(t)\kt+
	\int_{-\infty}^\infty dx\: v(x)\,|\psi(x,t)|^2\Big]\nn\\
	&=\frac{2}{\hbar}\int_{-\infty}^\infty dx\:\IM[v(x)]\,|\psi(x,t)|^2,
	\label{id-1}
	\end{align}	 
where ``$\RE$'' and ``$\IM$'' stand for the real and imaginary parts of their arguments, respectively.	
	\item We use \eqref{sch-eq-x} to show that
	\begin{align}
	\partial_t|\psi(x,t)|^2&=2\RE\big[\psi(x,t)^*\partial_t\psi(x,t)\big]\nn\\
	&=\frac{2}{\hbar}\IM\Big[\psi(x,t)^*\big\{-\frac{\hbar^2}{2m}\partial_x^2\psi(x,t)+v(x,t)\psi(x,t)\big\}\Big]\nn\\
	&=-\frac{\hbar}{m}\,\IM\Big[\psi(x,t)^*\partial_x^2\psi(x,t)\Big]+
	\frac{2}{\hbar}\,\IM\Big[v(x)\,|\psi(x,t)|^2\Big]\nn\\
	&=-\frac{\hbar}{m}\,\partial_x\IM\Big[\psi(x,t)^*\partial_x\psi(x,t)\Big]+
	\frac{2}{\hbar}\,\IM[v(x)]\,|\psi(x,t)|^2.
	\label{id-2}
	\end{align}
	
	\item We differentiate both side of \eqref{rho=} with respect to $t$ and use \eqref{id-1} and \eqref{id-2} to establish
	\begin{align}
	\partial_t\varrho(x,t)+\partial_x J(x,t)&=\Big\{\frac{2}{\hbar}\,\IM[v(x)]-\frac{d}{dt}\ln\big[\parallel\!\psi(t)\!\parallel^2\big]\Big\}\varrho(x,t).
	\label{CE}
	\end{align}
where 
	\be
	J(x,t):=
	\frac{\hbar\, \IM[\psi(x,t)^*\partial_x\psi(x,t)]}{m \,\parallel\psi(t)\parallel^2}.
	\ee

	\end{enumerate}
 
When $\widehat H$ is Hermitian, $v$ is necessarily real-valued, $\parallel\!\psi(t)\!\parallel$ is a constant, and $J$ is determined by the value of the position wave function and its derivative with respect to $x$ at $(x,t)$. Then, $\IM[v(x)]=0$ and \eqref{CE} reduces to the continuity equation describing the local conservation of the probability of the localization of the particle in the position space. When $v$ is not real-valued this local conservation law does not apply. Nevertheless, as we showed above, the total probability is conserved. This is consistent with the fact that \eqref{rho=} implies
	\be
	\int_{-\infty}^\infty dx\:\varrho(x,t)=1.
	\label{tot-conserve}
	\ee
We can also use \eqref{id-1} and  \eqref{CE} to show that the time derivative of the left-hand side of \eqref{tot-conserve} vanishes.

\section{Another argument for unitarity that uses the Born rule}

An important consequence of the general expression for the Born rule \eqref{BR-2} is that linear operators $\widehat O$ describing the physical observables of a quantum system must be Hermitian. To see this, first we recall that the expectation value of $\widehat O$ in a state $\sR$ of the system is the statistical average of outcomes of measurements of $\widehat O$ when the system is in this state. We can use  the Born rule \eqref{BR-2} to identify it with
	\be 
	\frac{\br\psi|\widehat O\psi\kt}{\br\psi|\psi\kt},
	\label{ex-va}
	\ee
where $\psi$ is an arbitrary state vector belonging to $\sR$. Because the outcomes of measurements of observable are the readings of a measuring device, its expectation values are real numbers. Most textbooks on quantum mechanics show that if a linear operator is Hermitian, then its expectation values are real \cite{sakurai,cohen-tannoudji}. They fail to even mention that the converse is also true, i.e., the reality of all expectation values implies the Hermiticity of the corresponding operator.\footnote{In Appendix~B we provide a simple proof of this statement for cases where $\sH$ is finite-dimensional. For a proof that applies also for the infinite-dimensional Hilbert spaces, see Ref.~\cite{Schechter}.} This means that if $\widehat O$ is non-Hermitian, there is always a state vector in which the expectation value of $\widehat O$ is not a real number. This disqualifies non-Hermitian operators to represent observables.\footnote{Because the calculation of expectation values uses the inner product on $\sH$, it is possible for some non-Hermitian operators to become Hermitian upon a proper modification of the inner product on $\sH$ which actually means a redefinition of $\sH$ as the Hilbert space associated with the quantum system 
\cite{review,bender}.} 

This observation provides an strong argument for the Hermiticity of the Hamiltonian operator, and consequently the unitarity of time evolution, if we follow the standard practice of identifying the Hamiltonian operator with the energy observable \cite{Dirac-QM}.  

It is important to realize that this assumption is not a part or consequence of any of the postulates of quantum mechanics. As we show below, it actually conflicts with the principle that the meaning and properties of physical quantities associated with a given system must be invariant under canonical transformations. 

Consider a time-dependent unitary operator $\widehat\cU(t)$ acting in the Hilbert space $\sH$ of a generic quantum system. Suppose that under the transformation, 
	\be
	\psi(t)\to\tilde\psi(t):=\widehat\cU(t)\psi(t),
	\label{trans1}
	\ee
the observables $\widehat O$ and the Hamiltonian $\widehat H(t)$ respectively transform to the linear operators $\widehat{\tilde O}$ and $\widehat{\tilde H}(t)$ such that the following two conditions hold.
	\begin{enumerate}
	\item The expectation value of $\widehat O$ computed for the state of the system given by $\psi(t)$ coincides with the expectation value of $\widehat{\tilde O}$ computed for the state of the system given by $\tilde\psi(t)$, i.e.,
	\be
	\frac{\br\psi(t)|\widehat O\psi(t)\kt}{\br\psi(t)|\psi(t)\kt}=
	\frac{\br\tilde\psi(t)|\widehat{\tilde O}\tilde\psi(t)\kt}{\br\tilde\psi(t)|\tilde\psi(t)\kt}.
	\ee
	\item $\tilde\psi(t)$ solves the Schr\"odinger equation,
	\be
	i\hbar\frac{d}{dt} \tilde\psi(t)=\widehat{\tilde H}(t)\tilde\psi(t).
	\label{sch-eq-tilde}
	\ee
	\end{enumerate}
Then, because $\widehat\cU(t)$ is a unitary operator, \eqref{sch-eq} and \eqref{trans1} -- \eqref{sch-eq-tilde} imply
	\begin{align}
	&\widehat{\tilde O}= \widehat\cU(t)\widehat O\,\widehat\cU(t)^{-1},
	\label{trans2}\\
	&\widehat{\tilde H}= \widehat\cU(t)\widehat H(t)\,\widehat\cU(t)^{-1}+
	i\hbar\Big[\frac{d}{dt}\,\widehat\cU(t)\Big]\,\widehat\cU(t)^{-1}.
	\label{trans3}
	\end{align} 
	
Unitary transformations of the form \eqref{trans1} accompanied with $\widehat O\to\widehat{\tilde O}$ and $\widehat H(t)\to\widehat{\tilde H}(t)$ are the quantum mechanical analogs of the classical canonical transformations \cite{Fock}. They provide equivalent descriptions of quantum systems.\footnote{This is because every physical quantity that we can compute in quantum mechanics is the expectation value of some linear operator.} In particular, the meaning of a physical quantity and its value obtained before and after performing such a transformation must coincide. 

According to \eqref{trans2} and \eqref{trans3}, the Hamiltonian operator does not transform similarly to the observables. In particular, its spectrum can change dramatically under a time-dependent quantum canonical transformation. An extreme example is when we identify $\widehat\cU(t)$ with the inverse of the evolution operator, i.e., set $\widehat\cU(t)=\widehat U(t,t_0)^{-1}$. Substituting this equation in \eqref{trans3}, we find $\widehat{\tilde H}(t)=\widehat 0$, where $\widehat 0$ is the zero operator. This in particular implies that the spectrum of the transformed Hamiltonian coincides with $\{0\}$. Therefore, even if we assume that $\widehat H(t)$ corresponds to the energy observable before this transformation, we cannot identify it with the energy observable following the transformation. This argument suggests that we should distinguish between the Hamiltonian operator and the Hermitian operator representing the energy observable. This distinction becomes particularly evident in the study of quantum systems with dynamical Hilbert spaces  \cite{prd-2018,entropy-2020}.

\section{Conclusion}

The original formulation of the postulates of quantum mechanics demands that the time evolution of state vectors be realized through unitary transformations of the Hilbert space. The standard expression for the Born rule is valid provided that the system abides by this assumption. Therefore any attempt to identify unitarity as a consequence of this expression is futile.

For the special class of systems described by standard Hamiltonian operators, the Hermiticity of the Hamiltonian, which is equivalent to the unitarity of the corresponding evolution operator, is a necessary condition for the local conservation of the probabilities of localization of the particle in the position space. This conservation law does not follow from the postulates of quantum mechanics and is highly sensitive to the structure of the Hamiltonian operator. For example, it fails to apply if we replace the kinetic energy term in the Hamiltonian with a non-polynomial function of $\widehat P$. Similarly, it does not apply for the localization of the particle in the momentum space unless the potential is a polynomial. These considerations suggest that the violation of local conservation of probabilities is not a strong argument against considering non-Hermitian Hamiltonians and non-unitary evolutions.

A stronger argument against non-Hermitian Hamiltonians is the requirement that  observables must be represented by Hermitian operators, which prevents non-Hermitian Hamiltonians to represent energy observables. The transformation property of the Hamiltonian operator under time-dependent canonical transformations suggests that we must distinguish between the Hamiltonian operator and the energy operator, because the latter has a different canonical transformation property. 
 
If we agree not to demand the Hamiltonian operator to represent an observable, we cannot use the measurement postulate to show that it must be a Hermitian operator.\\[6pt]

\noindent {\bf Conflict of interest statement:} The author declares that there are no conflicts of interest regarding the publication of this article.\\[6pt]	  
		  
\noindent {\bf Acknowledgements:} 
The author wishes to thank Alireza Shariati Joni for his comments and suggestions on the first draft of this article. This work has been supported by Turkish Academy of Sciences (T\"UBA).

 {
\subsection*{Appendix A: Linearity property of the Born rule \eqref{BR-15-t}}}

Consider the mixed state given by a density operator of the form
	\be
	\widehat\rho=c_1\,\widehat\rho_1+c_1\,\widehat\rho_2,
	\label{app-B1}
	\ee
where $c_1$ and $c_2$ are real and positive numbers, and $\widehat\rho_1$ and $\widehat\rho_2$ are density operators. Let $\fR$, $\fR_1$, and $\fR_2$ be the states of the ensemble of quantum states determined by the density operators $\widehat\rho$, $\widehat\rho_1$, and $\widehat\rho_2$.

If $c_1+c_2=1$, and $\widehat\rho_1$ and $\widehat\rho_2$ are normalized, i.e., have unit trace, then the same holds for $\rho$, and the Born rule \eqref{BR-15-t} gives
	\begin{align}
	{\rm Prob}_n&=
	{\rm tr}[\widehat \Pi_n\widehat{ \rho}]=
	c_1\,{\rm tr}[\widehat \Pi_n\widehat{\rho_1}]+
	c_2\,{\rm tr}[\widehat \Pi_n\widehat{ \rho_2}]\nn\\
	&=c_1\,{\rm Prob}_n^{(1)}+c_2\,{\rm Prob}_n^{(2)},
	\label{linearity}
	\end{align}
where
	\be
	{\rm Prob}_n^{(i)}:={\rm tr}[\widehat \Pi_n\widehat{\rho_i}],
	\label{appB-0}
	\ee
is the probability of finding $\omega_n$ for a measurement of $\widehat O$ when the system is in the state given by $\widehat{\rho_i}$, for $i\in\{1,2\}$. In this case, $c_1$ and $c_2$ respectively correspond to the probability of the ensemble to be in the states $\fR_1$ and $\fR_2$, which justifies \eqref{appB-0}.

Since under unitary time evolutions, normalized density operators remain normalized, the linearity property \eqref{linearity} of ${\rm Prob}_n$ applies for mixed states undergoing unitary evolutions. In the following we derive the analog of this property for general non-normalized density operators.

 {Suppose that $\widehat\rho_1$ and $\widehat\rho_2$ are arbitrary density operators, and $c_1$ and $c_2$ are arbitrary positive real numbers. Then, substituting \eqref{app-B1} in \eqref{BR-15-t}, we find
	\begin{align}
	{\rm Prob}_n&=\frac{{\rm tr}\big
	(c_1\widehat\Pi_n\widehat\rho_1+c_2\widehat\Pi_n\widehat\rho_1\big)}{
	{\rm tr}\big(c_1\,\widehat\rho_1+c_2\,\widehat\rho_1\big)}\nn\\
	&=\frac{c_1\,{\rm tr}\big(\widehat\rho_1\big)\,{\rm tr}\big(\widehat\Pi_n\widehat{\hat\rho_1}\big)
	+c_2\,{\rm tr}\big(\widehat\rho_2\big)\,{\rm tr}\big(\widehat\Pi_n\widehat{\hat\rho_2}\big)}{
	c_1 {\rm tr}\big(\widehat\rho_1\big)+c_2 {\rm tr}\big(\widehat\rho_1\big)}\nn\\
	&=\hat c_1{\rm tr}\big(\widehat\Pi_n\widehat{\hat\rho_1}\big)+
	\hat c_2{\rm tr}\big(\widehat\Pi_n\widehat{\hat\rho_2}\big),\nn\\[6pt]
	&=\hat c_1\,{\rm Prob}_n^{(1)}+\hat c_2\,{\rm Prob}_n^{(2)}
	\label{appB-2}	
	\end{align}
where	
	\begin{align}
	&\widehat{\hat\rho_i}:={\rm tr}\big(\widehat\rho_i\big)^{-1}\widehat\rho_i,
	\label{appB-3}\\
	&\hat c_i:=\frac{c_i\,{\rm tr}\big(\widehat\rho_i\big)}{
	c_1 {\rm tr}\big(\widehat\rho_1\big)+c_2 {\rm tr}\big(\widehat\rho_2\big)},
	\label{appB-4}
	\end{align}}

 {If we solve \eqref{appB-3} for $\widehat\rho_i$, substitute the result in \eqref{app-B1}, and use \eqref{appB-4}, we obtain
	\begin{align}
	\widehat\rho&=c_1\,{\rm tr}\big(\widehat\rho_1\big)\widehat{\hat\rho_1}+c_2\,{\rm tr}\big(\widehat\rho_2\big)\widehat{\hat\rho_2}\nn\\
	&=\big[c_1 {\rm tr}\big(\widehat\rho_1\big)+c_2 {\rm tr}\big(\widehat\rho_1\big)\big]
	\big[\hat c_1\widehat{\hat\rho_1}+\hat c_2\widehat{\hat\rho_2}\big]\nn\\
	&={\rm tr}\big(\widehat\rho)\big(\hat c_1\widehat{\hat\rho_1}+\hat c_2\widehat{\hat\rho_2}\big).
	\label{appB-5}
	\end{align}
We can use this equation, to express the normalized density operator \eqref{nor-rho} in the form
	\be
	\widehat{\hat\rho}=\hat c_1\widehat{\hat\rho_1}+\hat c_2\widehat{\hat\rho_2}.
	\label{appB-6}
	\ee
Because $\widehat{\hat\rho}$ has unit trace and describes the same mixed state as ${\widehat\rho}$,
	\eqref{BR-15-t} implies
	\be
	{\rm Prob}_n={\rm tr}\big(\widehat\Pi_n\widehat{\hat\rho}\big).
	\label{appB-last}
	\ee
It is important to realize that it is the coefficients $\hat c_1$ and $\hat c_2$ that respectively give the probability of the ensemble to be in the states $\fR_1$ and $\fR_2$.  
	
Equations \eqref{appB-2}, \eqref{appB-6}, and \eqref{appB-last} confirm the linearity property of ${\rm Prob_n}$ computed using \eqref{BR-15-t}.}\\[6pt]

\subsection*{Appendix B: Hermiticity as a consequence of the reality of the\linebreak expectation values}

Consider a linear operator $\widehat O$ acting in a finite-dimensional Hilbert space $\sH$. We wish to show that if all the expectation values of $\widehat O$ are real, then $\widehat O$ must be a Hermitian operator.

Let $\widehat K$ and $\widehat L$ be the Hermitian operators defined by: $\widehat K:=\frac{1}{2}(\widehat O+\widehat O^\dagger)$ and
$\widehat L:=\frac{1}{2i}(\widehat O-\widehat O^\dagger)$, so that 
	\be
	\widehat O=\widehat K+i \widehat L.
	\label{O=KL}
	\ee
Because $\widehat L$ is Hermitian, there is an orthonormal basis $\{\phi_n\}_{n=0}^N$ of $\sH$ consisting of the eigenvectors of $\widehat L$, i.e.,
	there are eigenvalues $\lambda_n$ such that $\widehat L\phi_n=
	\lambda_n\phi_n$. Because $\phi_n$'s are unit vectors, this equation implies
		\be
		\lambda_n=\br\phi_n|\widehat L\phi_n\kt.
		\label{eigenvalue}
		\ee
Furthermore, in view of the completeness relation for $\phi_n$'s, i.e.,
$\sum_{n=1}^N|\phi_n\kt\br\phi_n|=\widehat I$, we have
	\be
	\widehat L=\sum_{n=1}^N \lambda_n|\phi_n\kt\br\phi_n|.
	\label{complete-7}
	\ee
This equation implies that for every state vector $\psi\in\sH$,
	\be
	\br\psi|\widehat L\psi\kt=\sum_{n=1}^N \lambda_n\br\psi|\phi_n\kt\br\phi_n|\psi\kt=
	\sum_{n=1}^N \lambda_n|\br\phi_n|\psi\kt|^2.
	\label{ex-va-L}
	\ee
Because $\widehat L$ is Hermitian,
	\[\br\psi|\widehat L\psi\kt^*=
\br \widehat L\psi|\psi\kt=\br\psi|\widehat L^\dagger\psi\kt=\br\psi|\widehat L\psi\kt.\]
Therefore, $\br\psi|\widehat L\psi\kt$ is real. Similarly, the Hermiticity of $\widehat K$ implies that $\br\psi|\widehat K\psi\kt$ is real. 
	
Now, suppose that for every state vector $\psi\in\sH$, the expectation value of $\widehat O$ in the state given by $\psi$ is real. Then because $\br\psi|\widehat L\psi\kt$ and $\br\psi|\widehat K\psi\kt$ are both real and, in view of \eqref{O=KL},
	\be 
	\frac{\br\psi|\widehat O\psi\kt}{\br\psi|\psi\kt}=
	\frac{\br\psi|(\widehat K+i\widehat L)\psi\kt}{\br\psi|\psi\kt}=
	\frac{\br\psi|\widehat K\psi\kt}{\br\psi|\psi\kt}+i
	\frac{\br\psi|\widehat L\psi\kt}{\br\psi|\psi\kt},
	\nn
	\ee
this condition implies that $\br\psi|\widehat L\psi\kt=0$. Since this relation holds for every nonzero element of $\sH$, it must hold for the eigenvectors $\phi_n$ of $\widehat L$, i.e., $\br\phi_n|\widehat L\phi_n\kt=0$. In light of \eqref{eigenvalue}, this shows that $\lambda_n=0$. Substituting this in \eqref{complete-7}, we find $\widehat L=\widehat 0$. Therefore, according to 
\eqref{O=KL}, $\widehat O=\widehat K$. Because $\widehat K$ is a Hermitian operator, this proves the Hermiticity of $\widehat O$.

\ed